\documentclass[aps,amsmath,amssymb,twocolumn,preprintnumbers,groupedaddress,floatfix,tightenlines,superscriptaddress,footinbib]{revtex4-2}

\usepackage{graphicx} 
\usepackage{pstricks}
\usepackage[all]{xy}
\usepackage{hyperref}
\newcommand{\be}{\begin{equation}}
\newcommand{\ee}{\end{equation}}
\newcommand{\mc}{\mathcal}
\newcommand{\K}{\mc{K}}
\newcommand{\mbb}{\mathbb}
\newcommand{\nc}{\newcommand}
\nc{\beq}{\begin{equation}}
\nc{\eeq}{\end{equation}}
\nc{\bea}{\begin{eqnarray}}
\nc{\eea}{\end{eqnarray}}
\begin{document}

%%%%%%%%%%%%%%%%%%%%%%%%%%%%%%%%%%%%%%%%%%%%%%%
%%%%%%%%%%%%%%%%%%%%%%%%%%%%%%%%%%%%%%%%%%%%%%%
%%%%%%%%%%%%%%%%%%%%%%%%%%%%%%%%%%%%%%%%%%%%%%%
%%%%%%%%%%%%%%%%%%%%%%%%%%%%%%%%%%%%%%%%%%%%%%%
%%%%%%%%%%%%%%%%%%%%%%%%%%%%%%%%%%%%%%%%%%%%%%%
%%%%%%%%%%%%%%%%%%%%%%%%%%%%%%%%%%%%%%%%%%%%%%%
%%%%%%%%%%%%%%%%%%%%%%%%%%%%%%%%%%%%%%%%%%%%%%%
%%%%%%%%%%%%%%%%%%%%%%%%%%%%%%%%%%%%%%%%%%%%%%%

\title{Self-mirror Large Volume Scenario with de Sitter}

\author{Rui Sun}
\email{sunrui@kias.re.kr}
\affiliation{Korea Institute for Advanced Study,\\
85 Hoegiro, Dongdaemun-Gu, Seoul 02455, Korea}

%%%%%%%%%%%%%%%%%%%%%%%%%%%%%%%%%%%%%%%%%%%%%%%
%%%%%%%%%%%%%%%%%%%%%%%%%%%%%%%%%%%%%%%%%%%%%%%
%%%%%%%%%%%%%%%%%%%%%%%%%%%%%%%%%%%%%%%%%%%%%%%
%%%%%%%%%%%%%%%%%%%%%%%%%%%%%%%%%%%%%%%%%%%%%%%
%%%%%%%%%%%%%%%%%%%%%%%%%%%%%%%%%%%%%%%%%%%%%%%
%%%%%%%%%%%%%%%%%%%%%%%%%%%%%%%%%%%%%%%%%%%%%%%
%%%%%%%%%%%%%%%%%%%%%%%%%%%%%%%%%%%%%%%%%%%%%%%
%%%%%%%%%%%%%%%%%%%%%%%%%%%%%%%%%%%%%%%%%%%%%%%

\begin{abstract}

The large volume scenario has been an important issue for flux compactifications with T-dual non-geometric fluxes.  
As one solution to this issue, to naturally embed duality in string compactification, we investigate in self-mirror Calabi-Yau flux compactification with large volume scenario visited. 
In particular, at the large volume limit, the non-perturbative terms contribute a special dominant uplift term in the order of $\mathcal{O}\left(\frac{1}{\mathcal{V}^2}\right)$, while the $\alpha'$-corrections are trivialized due to the self-mirror Calabi-Yau construction. 
These in total contribute
to effective scalar potential 
in the same order as from F-term $\frac{D W. DW}{\mathcal{V}^2}$, and essentially give rise to  de Sitter vacua allowed by swampland conjectures.

\end{abstract}

\maketitle

\section{Introduction}

Calabi-Yau compactifications have been explored intensively in string model building. In which, large volume scenario plays an important role
as it allows to study string model building with higher order $\alpha'$ corrections suppressed at the large volume limit \cite{Balasubramanian:2005zx, Conlon:2005ki}.
In particular, it has been shown 
that in Type IIB flux compactifications with Calabi-Yau manifolds of more complex structure than K\"ahler moduli, {\emph{i.e.,}} $h^{1,2}>h^{1,1}$, the leading perturbative and non-perturbative corrections are suppressed at large volume limit. 
The volume of the Calabi-Yau manifold is exponentially large, leading to a range of string scales from the Planck mass to the TeV scale, realising for the first time the large extra dimensions scenario in string theory. However, the current phenomenology study has been mostly explored for Calabi-Yaus with small hodge number $h^{1,1}$, \emph{i.e.,} with small number of K\"ahler moduli. 
For large number of K\"ahler moduli, it has been a challenge topic because of the complexity of Calabi-Yau volume in K\"ahler moduli space. In our study, we intend to investigate flux compactifications with large number of K\"ahler moduli and large volume scenario revisited with duality.

This is essentially motivated from string compactification with generalized fluxes including T-dual non-geometric fluxes.  
By turning on both geometric and T-dual non-geometric fluxes, special perspectives are observed in string phenomenology and cosmology, such as in \cite{Hull:2004in, Hull:2005hk, Shelton:2005cf, Reid-Edwards:2006lap, Shelton:2006fd, Ihl:2007ah, Marchesano:2007vw, Aldazabal:2006up, Grana:2006hr, Grana:2008yw,  Garcia-Etxebarria:2016ibz}. String compactification with duality and double field theory were also studied in \cite{Blumenhagen:2015lta, Blumenhagen:2015xpa}, with axion monodromy inflation and de Sitter vacua constructed when T-dual non-geometric fluxes introduced in flux compactification. 
However, the large volume scenario
with T-dual non-geometric fluxes become puzzling, as these T-dual fluxes are supposed to arise from a T-dual/mirror dual framework.
In the moduli space sense, this is supposed to be in a mirror dual moduli space described by the complex structure instead of the K\"ahler moduli. One would say while the standard K\"ahler moduli space is in the large volume scenario, the T-dual framework appear to be in the small volume regime \footnote{Here we thank Liam Macalister for raising up this problem of non-geometric large volume scenario.}. Therefore, the large volume scenario which incorporates the T-dual/mirror framework deserves further investigations. 
The aim of this letter is exactly to give one solution to this problem.

To study flux compactification with duality, self-mirror Calabi-Yau manifolds become good candidates as they naturally allows T-dual fluxes embedding in the same Calabi-Yau compactification. 
As self-mirror Calabi-Yau manifold is mirror dual to themselves, to reach the large volume and large complex structure limits in one Calabi-Yau compactification will not involve another mirror Calabi-Yau manifold but itself in flux compactification. 
In other words, self-mirror compactification allows us to study the large volume and large complex structure scenario in one specific Calabi-Yau compactification. And this is one approach to study flux compactification with duality in the special self-mirror large volume scenario.

In particular, self-mirror Schoen Calabi-Yau threefold with $h^{1,2}=h^{1,1}=19$ has been investigated for modular superpotential in heterotic string theory and F-theory over $d P \times P^1$ \cite{Curio:1997rn, Curio:1998bv} as a good candidate for self-mirror compactification. 
The Schoen threefold is constructed with fiber products of rational elliptic surfaces with section \cite{Schoen:1988, Braun:2007sn}, and its quotient with ${\mathbb{Z}}_3 \times {\mathbb{Z}}_3$ fundamental group were utilized in constructing heterotic standard model \cite{Braun:2004xv, Braun:2005zv, Braun:2005ux, Braun:2005nv, Donagi:2000zf, Ovrut:2018qog}.
%As it was shown in \cite{Braun:2007sn}, quotient Calabi-Yau threefold can be construct from Schoen Calabi-Yau threefold with quotient $X=\tilde{X}/(\mathbb{Z}_3\times \mathbb{Z}_3)$ in which the fundamental group is $\pi_1(X)=\mathbb{Z}_3\times \mathbb{Z}_3$. 
Other interesting self-mirror Calabi-Yau manifolds with smaller hodge numbers are also constructed with fundamental groups, such as in \cite{Candelas:2008wb, Braun:2007sn, Braun:2009qy}. 
In \cite{Lee:2019wij}, small volume regime with infinite distance limit from self-mirror Schoen threefold is studied with Emergent Strings. 
In type II string theory, self-mirror K3 surface also shows interesting properties with worldsheet construction of T-dual non-geometric backgrounds \cite{Hull:2017llx}.

In the following work, we will in particular analyse self-mirror large volume scenario which naturally allows flux compactification with T-duality manifested.
Employing the formalism and notations in the related work \cite{Sun:2024abc},
all analysis are carried out in Type IIB
string compactification.
The main new observation here is the presence of de Sitter vacua from self-mirror non-perturbative contribution to the effective scalar potential, instead of anti-de Sitter vacua in the standard large volume scenario.

\section{Self-mirror Calabi-Yau Geometry}

Firstly, let us introduce the geometry of self-mirror Calabi-Yau of Schoen type in detail, and then study the large volume scenario with self-mirror property.
As complete intersections, Schoen Calabi-Yau threefold can be constructed as fiber products of two del Pezzo surfaces fibered over $\mathbb P^1$. In which, the del Pezzo surfaces $d P_9$ can be defined as blow-up of $\mathbb P^2$ at $9$ points where all $9$ points are distinct with no Kodaira fibers collide \cite{Braun:2007sn, Schoen:1988, Braun:2004xv, Donagi:2000zf, Braun:2005nv, Braun:2005ux, Ovrut:2002jk, Braun:2005zv, Braun:2007xh, Braun:2007vy, Braun:2007tp}.
As a complete intersection with the ambient variety
$\mathbb P^2 \times \mathbb P^1 \times \mathbb P^2$ and coordinates $x_i, t_j, y_k$, with
\begin{equation}
\Big( [x_0:x_1:x_2],~ [t_0:t_1],~[y_0:y_1:y_2]\Big) \in 
\mathbb P^2 \times \mathbb P^1 \times \mathbb P^2,
\end{equation}
the defining equation of Schoen Calabi-Yau threefold can be written with the zero-set in two equations 
\bea \nonumber
	&\tilde P(x,t,y) =
	t_0 \tilde P_1\big(x_0, x_1, x_2\big) + 
	t_1 \tilde P_2\big(x_0, x_1, x_2\big)
	=  0 \, ,
	\\ 
	&\tilde R(x,t,y) = 
	t_1 \tilde R_1\big(y_0, y_1, y_2\big) + 
	t_0 \tilde R_2\big(y_0, y_1, y_2\big)  
	=  0 \, ,
\eea
with multi-degrees $(3,1,0)$ and $(0,1,3)$ respectively. 
The first Chern class of line bundles on Schoen Calabi-Yau threefold $X$ form a lattice of dimension $h^{1,1}\big(\tilde X \big)=19$ resulting in $19$ K\"ahler moduli.
In which, the ambient space can be dealt with toric methods is with $h^{1,1}\big(\mathbb P^2\times\mathbb P^1\times\mathbb P^2\big)=3$ providing $3$ number of ambient divisors while the rest $16$ divisors come from the non-ambient space 
that are not toric and cause difficulty to derive the triple intersection numbers for the total volume of Schoen Calabi-Yau manifold.
Alternatively, recall that the Schoen construction can be written as  
\bea \label{CY19}
CY^{(19,19)}={\tiny 
	\left[\begin{array}{c|cc}P^2&3&0\\P^1&1&1\\P^2&0&3\end{array}\right]}=
dP\times _{P^1_y}dP,
\eea
which can also be constructed from
$T^2 \times K3={\tiny 
	\left[\begin{array}{c|cc}P^2&3&0\\P^1&0&2\\P^2&0&3\end{array}\right]}$
via Voisin-Borcea involution\cite{Curio:1997rn}.
The Hodge diamond of Schoen Calabi-Yau threefold $\tilde X$ with equal number of K\"ahler and complex structure moduli reads
{\small 
\begin{equation}
h^{p,q}\big(\tilde X\big) = ~
\vcenter{\xymatrix@!0@=7mm@ur{
		1 &  0  &  0  & 1 \\
		0 &  19 &  19 & 0 \\
		0 &  19 &  19 & 0 \\
		1 &  0  &  0  & 1 
}}.
\end{equation}}
Under a $\mathbb Z_3 \times \mathbb Z_3$-quotient action, the self-mirror Schoen Calabi-Yau threefold $\tilde X$ can be reduced into a quotient Schoen Calabi-Yau threefold $X$ involving only the toric space with $h^{1,1}\big(\mathbb P^2\times\mathbb P^1\times\mathbb P^2\big)=3$ \cite{Braun:2004xv}. Namely, $3$ number of ambient divisors enter into the total volume of the quotient Schoen Calabi-Yau geometry whose Hodge diamond then reduces to
{\small \begin{equation}
 \label{eq:SchoenZ3Z3Hodge}
 h^{p,q}\big(X \big)
 =h^{p,q}\big(\tilde X \Big/ \big(\mathbb Z_3 \times \mathbb Z_3\big)\big)=
 \vcenter{\xymatrix@!0@=7mm@ur{
 		1 &  0 &  0 & 1 \\
 		0 & 3 & 3 & 0 \\
 		0 & 3 & 3 & 0 \\
 		1 &  0 &  0 & 1 
 }}
 \,.
 \end{equation}}
%Here we will consider both toric and non-toric divisors and explicitly derive the triple intersection numbers in both ambient and non-ambient space in the next section.

\section{Large Volume Scenario}
The total volume of quotient Schoen Calabi-Yau threefold involving the ambient space can be represented by
\bea\label{qvolume}
{\cal V}
&=& {1\over 6}\kappa_{ijk}~t^i t^j t^k \nonumber \\
%&=&{1\over 6}\left(9 t_{0}^{(1)}t_{0}^{(1)}t_{0}^{(2)}+9 t_{0}^{(1)}t_{0}^{(2)}t_{0}^{(2)}+54 t^{(0)} t_{0}^{(1)}t_{0}^{(2)}\right)\\
&=&{3\over 2}\,
 t_{0}^{(1)}t_{0}^{(1)}t_{0}^{(2)}
 + {3\over 2} \,t_{0}^{(1)}t_{0}^{(2)}t_{0}^{(2)}
 +9\, t^{(0)} t_{0}^{(1)}t_{0}^{(2)}
,
\eea 
where $t_{0}^{(1)}, t_{0}^{(2)}$ correspond to the fiber two-cycle volume, and $t^{0}$ is the base two-cycle volume. 
The corresponding four-cycle moduli $\tau_i$ can be derived accordingly  with 
\bea \nonumber
&\tau_1 = \partial_{t_{0}^{(1)}} {\cal V} = {3\over 2}\, (t_{0}^{(2)})^2+ {3\over 2}\,t_{0}^{(1)} t_{0}^{(2)} + 9\, t^{(0)} t_{0}^{(2)} ,\\ \nonumber
&\tau_2 = \partial_{t_{0}^{(2)}} {\cal V} = {3\over 2}\, (t_{0}^{(1)})^2+ {3\over 2}\,t_{0}^{(1)} t_{0}^{(2)} + 9\, t^{(0)} t_{0}^{(1)} ,\\ 
&\tau_s = \partial_{t^{(0)}} {\cal V} = 9\, t_{0}^{(1)} t_{0}^{(2)} .
\eea
As follows,
\bea \nonumber
&t_{0}^{(1)} = \frac{\sqrt{\tau_0} \sqrt{6 \tau_2-\tau_0}}{3 \sqrt{6 \tau_1-\tau_0}},\\ \nonumber
&t_{0}^{(2)} = \frac{\sqrt{\tau_0} \sqrt{6 \tau_1-\tau_0}}{\sqrt{54 \tau_2-9 \tau_0}},\\ 
&t^{(0)} = \frac{-4 \tau_0 (\tau_1+\tau_2)+12 \tau_1 \tau_2+\tau_0^2}{6 \sqrt{\tau_0} \sqrt{6 \tau_1-\tau_0} \sqrt{6 \tau_2-\tau_0}},
\eea
and the total volume can be expressed with the k\"ahler moduli, such that
\bea\label{vQSchoen}
  {\cal V}= {1\over 18}\sqrt{\tau_0 (\tau_0-6 \tau_1)(\tau_0-6 \tau_2)}.
\eea
The total volume of quotient Schoen Calabi-Yau threefold is able to reach a large volume limit with divisor $\tau_0 \to \infty$ while $\tau_1$, $\tau_2$ being finite.
According to one single ambient fiber divisor $\tau_1$ or $\tau_2$, a bounded behavior shows for the volume of quotient Schoen Calabi-Yau manifold.
However, with both $\tau_1$ and $\tau_2$ goes to exponentially large, the total volume is also able to reach the large volume limit  with finite value of base divisor $\tau_0$. And a symmetric shape of the volume appears because of the double elliptic construction.
Being able to reach the large volume limit properly, 
quotient Schoen threefold allows us to study the effective scalar potential in self-mirror large volume scenario. 

In the standard large volume scenario, the $\alpha'$-correction to the effective scalar potential of type IIB string theory corresponds to the imaginary contribution $i \gamma \bigl( X^0\bigr)^2$ with $\gamma=\frac{\zeta(3)~\chi(\hat X)}{(2\pi)^3}$ \cite{Balasubramanian:2005zx, Conlon:2005ki}. While for self-mirror Calabi-Yau $X$, the imaginary contribution becomes trivial with vinishing Euler characteristic $\chi(\hat X)=0$. Therefore, the $\alpha'$-correction vanish trivially.
For self-mirror Calabi-Yaus, since the Euler characteristic $\chi(\hat X)=0$, the prepotential from self-mirror compactification reduces to
\be
\label{Fself}
%  F = {1\over 6}\kappa_{ijk} X^i X^j X^k/X^0 + \frac{1}{2} a_{ij} X^i X^j + b_i X^i X^0 + F_{\rm inst.}\,,
F = {1\over 6}\kappa_{ijk} X^i X^j X^k/X^0 + \frac{1}{2} a_{ij} X^i X^j + b_i X^i + F_{\rm inst.},
\ee
with $a_{ij}= \frac{1}{2} \int_{\hat X} \iota _*(c_1(J_j)) \wedge J_i~\textit{mode}~\mathbb{Z}$, and $b_{j}=\frac{1}{4!} \int_{\hat X} c_2(\hat X) \wedge J_j$ with $c_2(\hat X)$ as the second Chern. 
The K\"ahler potential can be represented in terms of the prepotential by
\bea
{\K} = -\ln[X^i \bar F_i(\bar X) + \bar X^i F_i(X)]\,.
\eea
The K\"ahler deformations can be purely captured by prepotential perturbatively from the terms with constant coefficients $a_{ij}, b_j$.
The instantons and/or gaugino condensation contributes non-perturbatively to the superpotential as well \cite{Witten:1996bn}. 
These contributions get incorporated into the effective scalar potential as in the $\mc{N}=1$ supergravity F-term scalar potential, resulting from the flux induced superpotential and K\"ahler potential,
\bea
\label{Vscalar}
V =  e^{\K} \left[G^{i \bar{j}} D_i W {D}_{\jmath} \bar{W} - 3 \vert W \vert^2 \right],
\eea
where $i,j$ running over all moduli, and $D_i W = \partial_i W + (\partial_i K) W,~ D_i \bar{W} = \partial_i \bar{W} + (\partial_i K) \bar{W}$. 
The no-scale scalar potential reduces to  
\bea\label{Vnoscale}
V_{no-scale} = e^{\K}~ G^{a\bar b}D_a W {D}_b \bar{W},
\eea
in which $a$ and $b$ depend on the dilaton and complex-structure moduli under the no-scale constraints.
The minimum of the $V_{no-scale}$ can be solved by 
\bea
D_a W \equiv \partial_a W + (\partial_a {\K}) W = 0,
\eea
while the complex structure and the dilaton moduli are stabilized by the introduced background fluxes. 
The value of $W$ at the vacuum can be denoted as $W_0$.

\section{Self-mirror Effective Scalar Potential}

The flux induced superpotential, {\emph{i.e.,}} the Gukov-Vafa-Witten\,(GVW) superpotential of $\mc{N}=1$ supergravity theory  results in \cite{Gukov:1999ya}
\bea
\label{GVW}
W = \int_X G_3 \wedge \Omega,
\eea
where $\Omega$ denotes the holomorphic $\left(3,0\right)$ form, $G_3 = \mathfrak F_3 - i S H_3$ contains the contribution of both the RR and NS three-form 
%\footnote{Note that for generalised fluxes, $H$-flux will be extended with T-dual $F, Q, R$-fluxes as well.}
while the chiral axio-dilaton $S$ represents
\bea
S =  e^{-\phi}-i C_0.
\eea
Here the superpotential depends not only on the dilaton, but also the complex structure
moduli measuring the size of these 3-cycles.

The K\"ahler potential $\mc K$ contains the contributions from all the three moduli: axion-dilaton, complex structure, and the K\"ahler moduli 
$\mc{K}= \mc{K}_{T} + \mc{K}_{U} + \mc{K}_{S}$
taking the form of
\bea
\mc{K} =
-2 \ln \left[\mc{V} \right] -
\ln\left[-i \int_M \Omega \wedge \bar{\Omega}\right] - \ln\left(S + \bar{S}\right),
\eea
in which $\cal V$ denotes the total volume of the Calabi-Yau manifold in the units $l_s = 2
\pi \sqrt{\alpha'}$.
The K\"ahler potential $\K_{T_i}$
 of no-scale type are constraint with $G^{i\bar{\jmath}} \K_i \K_{\bar{\jmath}} = 3$ in which the  K\"ahler metric is denoted by
$G_{i\bar{\jmath}} \equiv \mc{K}_{i\bar{\jmath}}=
\partial^2 \mc{K}/\partial_i  \partial_{\bar{\jmath}} $ and  $G^{i\bar{\jmath}}=G^{-1}_{i\bar{\jmath}}$.
With the dilaton and complex structure moduli integrated out, the K\"ahler potential for self mirror compactification with trivial Euler characteristic, reduces to 
\bea \label{Ksim}
{\K}  =  {\K}_{cs}
-2 \ln  \mc{V}.
\eea
Note that here the term corresponding to the $\alpha'$-correction in high order of K\"ahler potential $\frac{\xi}{2 g_s^{3/2}}$ is factored with $\xi = - \frac{\zeta\left(3\right) \chi\left(X\right)}{2(2 \pi)^3}$ that trivially vanishes for self-mirror Calabi-Yau manifolds with $\chi=0$. And thus this term does not appear in self-mirror effective potential.

The superpotential contributes to the scalar potential non-perturbatively 
with D3-brane instantons and/or gaugino condensation when wrapped D7-branes introduced \cite{Witten:1996bn}, such that 
\bea
\label{supercorrect}
W = \int_M G_3 \wedge \Omega + \sum_i A_i e^{- a_i T_i},
\eea
where $A_i$ represents the one-loop determinant depending on the complex structure moduli,
and $a_i = \frac{2 \pi}{K}$ with $K=1$ for D3-instanton case, $K \in \mbb{Z}_+$ for general cases. 
And the complexified K\"ahler moduli are denoted by $T_i \equiv \tau_i +i b_i$. 
Set $\tau_1, \tau_2, \tau_0$ correspond to the divisors $D_1$,  $D_2$ and $D_0$ which can be embeded into a Calabi-Yau 4-fold with elliptic torus fibration over a base threefold,
the non-perturbative contributions are manifested in the superpotential as  
\bea
W = W_0 + A_0 e^{- a_0 T_0}+ A_1 e^{-a_1 T_1} + A_2 e^{-a_2 T_2},
\eea
in which the dilaton and complex structure moduli dependences are integrated out to $W_0$.
Consider the K\"ahler potential \eqref{Ksim} with  trivial $\xi$ term (and thus trivial $\alpha'$ correction) reduces to 
\bea
\K =\K_0= -2 \ln {\cal V},
\eea
the non-perturbative terms from k\"ahler moduli contribute to the effective scalar potential \eqref{Vscalar} as 
\bea \label{VQ} \nonumber
&V& ={1\over \mc{V}^2} \left[ G^{i \bar{j}} \left(a_i A_i a_j \bar{A}_j e^{- \left(a_i T_i + a_j \bar{T}_j\right)} \right.\right.\\
&~&\left.\left. -( a_i A_i e^{-a_i T_i} \bar{W} \partial_{{j}} \K + a_j \bar{A}_j e^{- a_j \bar{T}_j} W \partial_{i} \K) \right) \right].
\eea
To obtain the non-perturbative terms, we first compute the inverse of K\"ahler metric that will be necessary as
\bea\label{Kmetric}
G^{i\bar{j}}=
\begin{pmatrix}
\tau_1^2-\frac{\tau_1 \tau_0}{3}+\frac{\tau_0^2}{18} & {\tau_0^2\over 36} & \qquad {\tau_0^2 \over 6}\\
\newline\\
{\tau_0^2 \over 36} & \tau_2^2-\frac{\tau_2 \tau_0}{3}+\frac{\tau_0^2}{18} & \qquad {\tau_0^2 \over 6}\\
\newline\\
{\tau_0^2 \over 6} & {\tau_0^2 \over 6} & \qquad \tau_0^2
\end{pmatrix}
\eea
and it can be verified that $G^{i\bar{j}}\partial_i {\cal K}~\partial_{\bar{j}} {\cal K}= 3$, with no-scale structure \eqref{Vnoscale} preserved.
%%%%%%%%%%%%%%%%%%%%%%%%%%%%%%%%
%%%%%%%%%%%%%%%%%%%%%%%%%%%%%%%%%%%%%%%

\subsection*{Large Volume Limit I}

\begin{figure}[h]
 	\centering 	\includegraphics[scale=0.42]{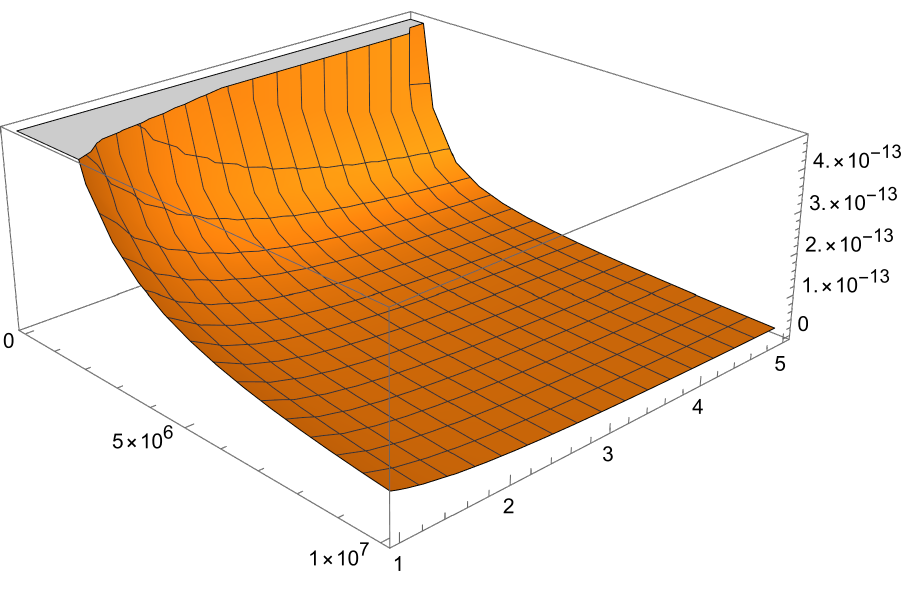}
\caption{Self-mirror scalar potential $V$ according to $\tau_0$ and $\tau_1$ with numerical constant value of 
$W_0=1, A_0=1,  a_0= 2 \pi, b_0= 1/4, \tau_1=10^7$. The scalar potential approaches positive value from above at the large volume limit, and therefore de Sitter vacuum approached.}
\label{figure:SPotential-QS01}
  \end{figure}

\begin{figure}[h]
 	\centering
\includegraphics[scale=0.55]{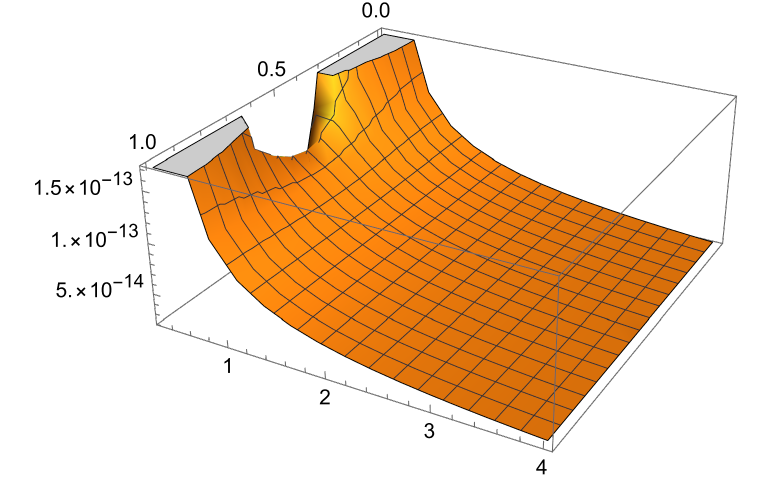}
\caption{Self-mirror scalar potential at the large volume limit according to $\tau_0$ and axionic field $b_0=[0, 1]$. Here the value $W_0=1,A_0=1,  a_0=2\pi, \tau_1= 10^7, \tau_2= 10^7$ are used.
The positive uplift term at $ a_0\tau_0\sim \ln \mc{V}$ provides  de Sitter vacuum around the saddle point region.}
\label{figure:SPotential-QS02}
\end{figure}

In the large volume limit I with fiber K\"ahler moduli $\tau_1, \tau_2$ both to be exponentially large and the base divisor $\tau_0$ with a finite value, we have the non-perturbative contribution as 
\bea
W = W_0 + A_0 e^{-a_0 T_{0}}.
\eea
The involved inverse of K\"ahler metric element can be read off from \eqref{Kmetric} with $G^{00}=\tau_0^2$. 
Note that the axionic field $b_i$ can be stabilized, we have $\partial_{{T_i}} \K=\partial_{{\tau_i}} \K:=\partial_{{i}} \K$,
while $b_i$ considered as constant coefficients.
Then the non-perturbative scalar potential results in
\bea \label{VSchoen0} \nonumber
V&=&\frac{W_0^2 (\tau_0 \tau_1 \tau_2)^2}{81\mc{V}^6} 
-\frac{ 2 W_0^2 \tau_0^3 \tau_1 \tau_2(\tau_1+\tau_2)}{243\mc{V}^{6}}\\
&~&+\frac{2 A_0 W_0 e^{-a_0 \tau_0} (\tau_0 \tau_1 \tau_2)^2\cos(a_0 b_0)(1+a_0 \tau_0)}{81\mc{V}^{6}}.
\eea
Consider that the total volume of quotient Schoen Calabi-Yau manifolds in the large volume limit behaves as
{\bea \label{vSchoen-q} \nonumber
  {\cal V} &=& 
  %{1\over 2}\sum_{i,j=0}^{8}\sqrt{\tau_0 (\tau_0+2 \tau_{i}^{(1)})(\tau_0+2 \tau_{j}^{(2)})}=
  {1\over 18}\sqrt{\tau_0^3-6 \tau_0^2(\tau_1+\tau_2)+ 36 \tau_0 \tau_1 \tau_2} \\ 
 &\sim & {1\over 3} \sqrt{ \tau_0\tau_1 \tau_2}, ~~
 \text{with}~ \tau_0 \tau_1 \tau_2 \sim (3 \mc{V})^{2},
\eea}
the non-perturbative effective scalar potential becomes
\be\label{VSchoen001}
V =\frac{W_0^2}{\mc{V}^2} 
-\frac{4 W_0^2 \tau_0^{3/2}}{9\mc{V}^{3}}
+\frac{2  A_0 W_0 e^{-a_0 \tau_0}\cos(a_0 b_0)(1+a_0 \tau_0)}{\mc{V}^{2}}.
\ee
Regard $V =V(\mc{V}, \tau_0)$, the minimum shall be solved at 
\be
\frac{\partial V} {\partial \mc{V}}
= \frac{\partial V} {\partial \tau_0}
= 0,
\ee
in which $\frac{\partial V} {\partial \tau_0}= 0$ can be %hyperbolically 
solved at 
\bea \label{tau}
 \mc{V} &=&-\frac{e^{a_0 \tau_0} W_0}{3 a_0^2 A_0 \cos(a_0 b_0) \sqrt{\tau_0} }, ~\text{with} \\ \nonumber
\frac{\partial V} {\partial \mc{V}} = 0
 &\Rightarrow &
e^{a_0 \tau_0} W_0 + 2A_0(1+a_0 \tau_0(1+a_0\tau_0)) \cos(a_0 b_0)=0
\eea
in which $a_0 \tau_0 \gg 1$ shall be required to suppress the higher order instanton corrections. Therefore, \eqref{tau} can be solved with 
\be
\tau_0=- \frac{2}{a_0} ProductLog\left[-\frac{1}{2} 
\sqrt{-\frac{W_0 }{2 A_0 \cos(a_0 b_0)}}\right]>0,
\ee
while $a_0>0, A_0>0, b_0>0$, and $ProductLog[z]$ denotes the principal solution for $w$ in $z=w e^w$.
Obviously, the large volume limit is approached while $a_0 b_0 \to \pi/2, 3\pi/2$.
The scalar potential under the decompactification limit $a_0 \tau_0 \sim \ln \mc{V}$
reduces to 
\be
V = \frac{W_0^2}{\mc{V}^2} 
-\frac{4 W_0^2 \tau_0^{3/2}}{9\mc{V}^{3}}
+\frac{2 A_0 W_0 \cos(a_0 b_0)(1+\ln \mc{V})}{\mc{V}^{3}},
\ee
in the order of $\mathcal{O} (\frac{1}{\mc{V}^{n}})$. 
In the large volume limit, the leading term $W_0^2 \over \mc{V}^2$ takes the dominant value which is obviously positive in the order of $\mathcal{O} (\frac{1}{\mc{V}^{2}})$, and the subleading terms suppressed at the order of $\mathcal{O} (\frac{1}{\mc{V}^{3}})$ and $\mathcal{O} (\frac{\ln \mc{V}}{\mc{V}^{3}})$ respectively. Therefore, in total, the non-perturbative scalar potential takes a \emph{positive} value while the complex structures and axio-dilatons at the vacuum integrated out to $W_0$. 
Intriguingly, the non-perturbative leading term contributes to the effective scalar potential in the same order as from F-term $\frac{D W. DW}{\mc{V}^2}$ in $\mathcal{O} (\frac{1}{\mc{V}^{2}})$ and essentially give rise to de Sitter vacua. 
As illustrated in Figure \ref{figure:SPotential-QS01} according to \eqref{VSchoen0}, the effective scalar potential approaches the positive uplift 
at large volume limit, and the resulting de Sitter minimum is again illustrated in Figure \ref{figure:SPotential-QS02}.

\subsection*{Large Volume Limit II}

In the large volume limit $\mc{V} \to \infty$ with $\tau_0 \to \infty$ exponentially, the non-perturbative superpotential terms approach
\be
W = W_0 + A_1 e^{-a_1 T_1} + A_2 e^{-a_2 T_2}.
\ee
The effective scalar potential \eqref{VQ}, with inverse of K\"ahler metric element $G^{i\bar{j}}$ ($i,j=1,2$) from \eqref{Kmetric}, become
{\small \bea \label{VQ2}
&V &= \frac{W_0^2 \tau_0^6}{17496~\mc{V}^6} 
-\frac{ W_0^2 (\tau_1+\tau_2)\tau_0^5}{2187~ \mc{V}^{6}}
\\
\nonumber
& &-\frac{\left(a_1 A_1 e^{-a_1 \tau_1} \cos(a_1 b_1) + a_2 A_2  e^{-a_2 \tau_2} \cos(a_2 b_2)\right)W_0 \tau_0^7}{104976 ~\mc{V}^{6}}  
\\
\nonumber
& & 
+\frac{\left(a_1^2 A_1^2 e^{-2 a_1 \tau_1}+ a_2^2 A_2^2 e^{-2 a_2 \tau_2}\right)\tau_0^8}{1889568~ \mc{V}^{6}}
\\
\nonumber
& & 
+\frac{\left(a_1 A_1 a_2 A_2 e^{-a_1 \tau_1-a_2 \tau_2}\cos(a_1 b_1- a_2 b_2)\right)\tau_0^8}{1889568~ \mc{V}^{6}}.
\eea}
The total volume of quotient Schoen Calabi-Yau \eqref{vQSchoen} in the large volume limit approaches 
\bea\nonumber
\mc{V}= {1\over 18}\sqrt{\tau_0 (\tau_0-6 \tau_1)(\tau_0-6 \tau_2)}\\
 ~\sim {1\over 18}~ \tau_0^{{3 \over 2}}, ~~
 \text{with}~ \tau_0 \sim (18\mc{V})^{2 \over 3},
\eea
and therefore
the effective scalar potential reduces to
{\small \bea \label{Vt12}
&V &= \frac{36 W_0^2}{\mc{V}^2} 
-\frac{ 8 (\frac{2}{3})^{1/3}W_0^2 (\tau_1+\tau_2)}{\mc{V}^{8/3}}
\\
\nonumber
& &-\frac{2^{2/3} 3^{4/3}\left(a_1 A_1 e^{-a_1 \tau_1} \cos(a_1 b_1) + a_2 A_2  e^{-a_2 \tau_2} \cos(a_2 b_2)\right)W_0 }{\mc{V}^{4/3}} 
\\
\nonumber
& & 
+\frac{2^{1/3} 3^{2/3}\left(a_1^2 A_1^2 e^{-2 a_1 \tau_1}+ a_2^2 A_2^2 e^{-2 a_2 \tau_2}\right)}{ \mc{V}^{2/3}}
\\
\nonumber
& & 
+\frac{2^{1/3} 3^{2/3}\left(a_1 A_1 a_2 A_2 e^{-a_1 \tau_1-a_2 \tau_2}\cos(a_1 b_1- a_2 b_2)\right)}{\mc{V}^{2/3}}.
\eea}
By taking the decompactification limit for fiber divisor 
\bea
a_1 \tau_1=\ln \mc{V},~ a_2 \tau_2=\ln \mc{V},
\eea
the effective scalar potential then reduces to
{\small \bea
&V &= \frac{36 W_0^2}{\mc{V}^2} 
-\frac{ 8 (\frac{2}{3})^{1/3}W_0^2 (\tau_1+\tau_2)}{\mc{V}^{8/3}}
\\
\nonumber
& &-\frac{2^{2/3} 3^{4/3}\left(a_1 A_1  \cos(a_1 b_1) + a_2 A_2   \cos(a_2 b_2)\right)W_0 }{\mc{V}^{7/3}} 
\\
\nonumber
& & 
+\frac{2^{1/3} 3^{2/3}\left(a_1^2 A_1^2 + a_2^2 A_2^2+a_1 A_1 a_2 A_2 \cos(a_1 b_1- a_2 b_2)\right)}{\mc{V}^{8/3}}.
\eea}

\begin{figure}[h]
 	\centering 	\includegraphics[scale=0.5]{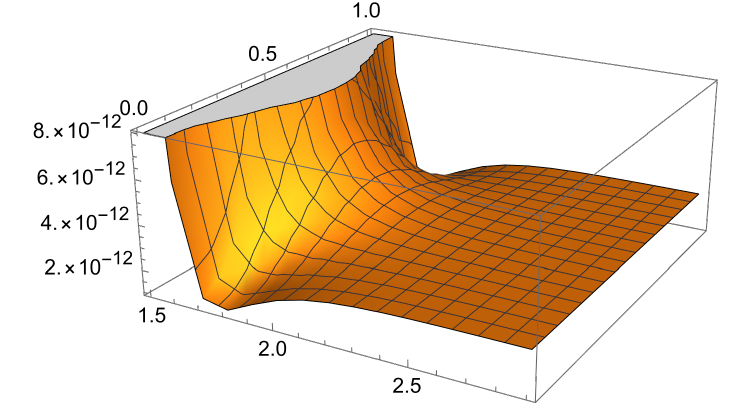}
 	\caption{Self-mirror scalar potential of quotient Schoen Calabi-Yau, at the large volume limit $\tau_0 \to \infty$, $a_1\tau_1, a_2\tau_2 \sim \ln \mc{V}$,   as function $\tau_2$ and axionic field $b_2=[0, 1]$ according to \eqref{VQ2}.
    Here the value $ W_0=1, A_1=1, A_2=1, a_1= 2\pi, a_2=2 \pi, \tau_0= 10^5, \tau_1= 2, b_1=1/2$ are used.
  The positive uplift term gives rise to de Sitter vacuum along the $\tau_1, \tau_2$ directions around the saddle point region as illustrated.
  }
 \label{figure:SPotential-QS03}
  \end{figure}

In the large volume limit, the leading term $36 W_0^2 \over \mc{V}^2$ takes the dominant value which is obviously positive in the order of $\mathcal{O} (\frac{1}{\mc{V}^{2}})$, while the subleading terms suppressed at the order of $\mathcal{O} (\frac{1}{\mc{V}^{7/3}})$ and $\mathcal{O} (\frac{1}{\mc{V}^{8/3}})$ respectively. Therefore, in total, the non-perturbative scalar potential contributes a positive value, while $W_0$ represents the value of the superpotential integrated out from the complex structures and axio-dilatons. 
Again regard the scalar potential \eqref{Vt12} as $V=V(\mc{V}, \tau_1, \tau_2)$, the possible minimum shall be solved at 
\be
\frac{\partial V} {\partial \mc{V}}
=\frac{\partial V} {\partial \tau_1}
= \frac{\partial V} {\partial \tau_2}=0.
\ee
From the symmetric construction of volume form with $\tau_1$ and $\tau_2$, one can solve $\mc{V}$ 
with the value of $\mathcal{V}^{2/3}$ results in
{\small\be \label{V23}
\frac{2^{1/3} 3^{2/3} W_0 (a_1^2 A_1 x \cos(a_1 b_1)- a_2^2 A_2  y\cos(a_2 b_2))}
{2 a_1^3 A_1^2 x^2 -2 a_2^3 A_2^2 y^2+(a_1-a_2) a_1 A_1 a_2 A_2 x y \cos(a_1 b_1-a_2 b_2)},
\ee}
where $x:=e^{-a_1\tau_1},~ y:=e^{-a_2\tau_2}$. Conveniently, by further taking the decompactification limit $a_1 A_1=W_0,\, a_2 A_2=W_0$,
the scalar potential results in the structure
\bea
&V& =\frac{36 W_0^2}{\mc{V}^2} -\frac{\kappa (\tau_1+\tau_2)}{\mc{V}^{8/3}}\\ \nonumber
& & - \frac{\lambda(e^{-a_1 \tau_1} \cos(a_1 b_1) + e^{-a_2 \tau_2} \cos(a_2 b_2))}{\mc{V}^{4/3}}
\\ \nonumber
& & +\frac{\mu (e^{-2 a_1 \tau_1}+ e^{-2 a_2 \tau_2}+ e^{- a_1 \tau_1- a_2 \tau_2}    \cos(a_1 b_1- a_2 b_2)}{\mc{V}^{2/3}},
\eea
while $\kappa= 8 (\frac{2}{3})^{1/3} W_0^2$, $\lambda=2^{2/3} 3^{4/3}W_0^2$, $\mu=2^{1/3} 3^{2/3}W_0^2$.
The value of $\tau_1, \tau_2$ may be further solved with the relation of $y$ and $x$,
\be
y=-\frac{\kappa +2 \mu a_1 x^2 \mc{V}^2- \lambda a_1  \cos(a_1 b_1) x\mc{V}^{4/3}}{\mu a_1 \cos(a_1b_1-a_2b_2)x \mc{V}^2},
\ee 
as illustrated in Figure \ref{figure:SPotential-QS03}, where the uplift term takes the dominant effect and essentially gives rise to de Sitter vacua with swampland conjectures criteria satisfied \cite{Sun:2024abc}.

\section{Conclusion}

The self-mirror large volume scenario from quotient Schoen Calabi-Yau compactification gave rise to a special uplift term to the effective scalar potential from non-perturbative contribution.
This uplift term is intriguingly positive in order $\mathcal{O} (\frac{1}{\mc{V}^{2}})$ and  essentially lead to de Sitter vacua.
In precise, these non-perturbative leading terms $W_0^2 \over \mc{V}^2$ and $36 W_0^2 \over \mathcal{V}^2$ became the dominant contribution from large volume scenario, while the contributions from complex structures and axio-dilaton are flux stabilized and integrated out to $W_0$ at the vacuum. 
In particular, the special uplift term is in the same order as from F-term $\frac{D W. DW}{\mathcal{V}^2}$  and KKLT mechanism 
in $\mathcal{O} (\frac{1}{\mathcal{V}^{2}})$. 
Considering that the special uplift term arises from self-mirror large volume scenario, we propose a new mechanism to obtain de Sitter vacuum.

%\begin{acknowledgements}
\emph{Acknowledgements:} 
We would in particular like to thank Ralph Blumenhagen, Fernando Quevedo, Jie Zhou for many helpful comments, and thank
Michele Cicoli, Wolfgang Lerche, Andre Lukas, Dieter L\"ust,  Pramod Shukla for helpful discussions.
We would also like to thank Xin Gao, Chuying Wang, Xin Wang, Shing Tung Yau and Piljin Yi for helpful discussions on related topics.
RS is supported by KIAS New Generation Research Grant PG080701, PG080704, and acknowledge Ludwig Maximilian University of Munich, Max Planck Institute of Physics for their hospitality, and LMU-China Academic Network for their support when this research was initialized.

%\end{acknowledgements}

%%%%%%%%%%%%%%%%%%%%%%%%%%%%%%%%%%%%%%%%%%%%%%%
%%%%%%%%%%%%%%%%%%%%%%%%%%%%%%%%%%%%%%%%%%%%%%%
%%%%%%%%%%%%%%%%%%%%%%%%%%%%%%%%%%%%%%%%%%%%%%%
%%%%%%%%%%%%%%%%%%%%%%%%%%%%%%%%%%%%%%%%%%%%%%%
%%%%%%%%%%%%%%%%%%%%%%%%%%%%%%%%%%%%%%%%%%%%%%%
%%%%%%%%%%%%%%%%%%%%%%%%%%%%%%%%%%%%%%%%%%%%%%%
%%%%%%%%%%%%%%%%%%%%%%%%%%%%%%%%%%%%%%%%%%%%%%%
%%%%%%%%%%%%%%%%%%%%%%%%%%%%%%%%%%%%%%%%%%%%%%%

\baselineskip=1.6pt

\bibliography{reference}

%%%%%%%%%%%%%%%%%%%%%%%%%%%%%%%%%%%%%%%%%%%%%%%
%%%%%%%%%%%%%%%%%%%%%%%%%%%%%%%%%%%%%%%%%%%%%%%
%%%%%%%%%%%%%%%%%%%%%%%%%%%%%%%%%%%%%%%%%%%%%%%
%%%%%%%%%%%%%%%%%%%%%%%%%%%%%%%%%%%%%%%%%%%%%%%
%%%%%%%%%%%%%%%%%%%%%%%%%%%%%%%%%%%%%%%%%%%%%%%
%%%%%%%%%%%%%%%%%%%%%%%%%%%%%%%%%%%%%%%%%%%%%%%
%%%%%%%%%%%%%%%%%%%%%%%%%%%%%%%%%%%%%%%%%%%%%%%
%%%%%%%%%%%%%%%%%%%%%%%%%%%%%%%%%%%%%%%%%%%%%%%
\end{document}